%% file: main.tex
 \documentclass[acmsmall]{acmart}
\AtBeginDocument{%
  }

\usepackage{multirow} 

\usepackage{xspace}
\usepackage{tcolorbox}
\usepackage{wrapfig} 
\usepackage{siunitx} 
\usepackage{amsmath}
\usepackage{dsfont}
\usepackage[misc]{ifsym}

\NewDocumentCommand{\answerbox}{ m +m }{%
    \begin{tcolorbox}[
        colback=white,
        colframe=gray!60,
        coltitle=black,
        title=#1, 
        fonttitle=\bfseries,
        boxrule=0.8pt,
        arc=0pt,
        top=4pt,
        bottom=4pt
        ]
        #2
    \end{tcolorbox}%
}

\newcommand{\approach}{{InlineCoder}\xspace}

\newcommand{\upperAnchor}{Draft\;}
\newcommand{\loweranchor}{draft\xspace}

\setcopyright{cc}
\setcctype{by-nc-nd}
\acmJournal{PACMSE}
\acmYear{2026} \acmVolume{3} \acmNumber{FSE} \acmArticle{FSE066}
\acmMonth{7} \acmDOI{10.1145/3797094}

\begin{document}

\title{In Line with Context: Repository-Level Code Generation via Context Inlining}


\author{Chao Hu}
\authornote{Both authors contributed equally to this research.}
\email{ythere@sjtu.edu.cn}
\affiliation{%
    \institution{Shanghai Jiao Tong University}
    \country{China}
}
\author{Wenhao Zeng}
\authornotemark[1] 
\email{zengwh_cs@sjtu.edu.cn}
\affiliation{%
    \institution{Shanghai Jiao Tong University}
    \country{China}
}
\author{Yuling Shi}
\email{yuling.shi@sjtu.edu.cn}
\affiliation{%
    \institution{Shanghai Jiao Tong University}
    \country{China}
}
\author{Beijun Shen}
\email{bjshen@sjtu.edu.cn}
\affiliation{%
    \institution{Shanghai Jiao Tong University}
    \country{China}
}
\author{Xiaodong Gu}
\authornote{Corresponding author.}
\email{xiaodong.gu@sjtu.edu.cn}
\affiliation{%
    \institution{Shanghai Jiao Tong University}
    \country{China}
}

\renewcommand{\shortauthors}{Hu et al.}

\begin{abstract}

    Repository-level code generation has attracted growing attention in recent years. Unlike function-level code generation, it requires the model to understand the entire repository and reason over complex dependencies across functions, classes, and modules. However, existing approaches such as retrieval-augmented generation (RAG) or context-based function selection often fall short; they primarily rely on surface-level similarity and struggle to capture the rich dependencies that govern repository-level semantics.
    In this paper, we introduce \approach, a novel framework for repository-level code generation. \approach enhances the understanding of repository context by inlining the unfinished function into its call graph, thereby reframing the challenge of repository understanding into a simpler function-level coding task. Given a function signature, \approach first generates a \textit{\loweranchor} completion (termed an "anchor"),
    which approximates downstream dependencies and enables perplexity-based confidence estimation. This anchor drives a bidirectional inlining process: (\emph{i}) Upstream Inlining, which embeds the anchor into its callers to capture diverse usage scenarios; and (\emph{ii}) Downstream Retrieval, which integrates the anchor's callees into the prompt to provide precise dependency context. The enriched context, combining draft completion with upstream and downstream perspectives, equips the LLM with a comprehensive repository view.
    Extensive experiments on the \textsc{DevEval} and \textsc{RepoExec} benchmarks demonstrate that \approach substantially outperforms a wide range of state-of-the-art baselines, achieving average relative gains of 29.73\% in EM, 20.82\% in ES, and 49.34\% in BLEU on \textsc{RepoExec} compared to the strongest baseline.
    These results highlight its effectiveness in repository context understanding as well as its generalization across domains.

\end{abstract}





\maketitle

\section{Introduction}


With the rapid development of code LLMs, \textit{repository-level code generation} has attracted growing attention in recent years~\cite{zhang2023repocoder, shrivastava2023repofusion, li2024deveval, le2024repoexec}. Unlike function-level generation, repository-level generation requires reasoning over entire repositories, accounting for coding conventions, API usage, and intricate inter-function dependencies~\cite{shrivastava2023repository}. Success in this setting demands not only syntactically correct and semantically valid code but also consistency with the repository’s broader design and dependencies~\cite{hassen2017scalable,le1996structural,maletic2001supporting,plotkin2004origins,zhao2015reusable}.

A key obstacle in repository-level generation is the repository context itself. While it contains the crucial information needed for accurate generation, directly feeding all files into an LLM is infeasible due to context window limitations and the overwhelming amount of irrelevant or redundant code~\cite{shi2023large}. This raises a central challenge: \textit{How can we distill and represent the most relevant context from a vast codebase to support effective code generation?}

Prior work has approached this challenge through various retrieval strategies. The most common is retrieval-augmented generation (RAG), which retrieves similar code snippets to the unfinished function~\cite{zhang2023repocoder,liang2024repofuse,su2024dragin,gao2023retrieval,gao2024preference, liu2024graphcoder}. However, similarity does not necessarily imply relevance in code: lexically similar snippets may be functionally unrelated, leading to noisy or misleading prompts. Agent-based pipelines further extend these capabilities by enabling iterative retrieval and LLM-guided evaluation of retrieval targets~\cite{bi2024iterative, ma2025alibaba, zhang2024codeagent}. More advanced methods incorporate program analysis, such as control-flow or data-flow graphs, to capture semantic dependencies more accurately~\cite{cheng2024dataflow, liu2024graphcoder, ouyang2024repograph, codexgraph, phan2025repohyper}. While effective for fine-grained tasks such as line-level completion or API prediction, these methods rely heavily on local context and often fail to generalize to function-level generation, where entire implementations must be synthesized from scratch.

To overcome these limitations, we propose \textbf{\approach}, a novel framework for repository-level code generation.
Our key insight is that a function's role within a repository is determined by its position in the repository's call stack: it is constrained by its \textit{upstream} callers (how it is used) while its implementation depends on its \textit{downstream} callees (what it depends on).
\approach enhances context understanding by inlining unfinished functions into their call graph, thereby reformulating repo-level generation into the more tractable function-level generation task.
Given a function signature, \approach first produces a draft implementation — an \textit{anchor} that approximates potential dependencies. This \loweranchor then drives a bidirectional inlining process: (1) \textbf{Upstream Inlining}, where we inline the anchor into its callers to provide rich usage scenarios. (2) \textbf{Downstream Retrieval}, where we incorporate all callees invoked by the \loweranchor, capturing its dependency context. Finally, we integrate the \loweranchor code with the inlined context into a comprehensive prompt, enabling the LLM to generate the final, contextually-aware code.

We evaluate \approach on two widely-used repository-level code generation benchmarks, \textsc{DevEval}~\cite{li2024deveval} and \textsc{RepoExec}~\cite{le2024repoexec}, using three backbone LLMs: \texttt{DeepSeek-V3}~\cite{liu2024deepseek}, \texttt{Qwen3-Coder}~\cite{yang2025qwen3}, and \texttt{GPT5-mini}~\cite{openai2025introducing}. \approach demonstrates the best overall performance. Notably, on \textsc{RepoExec}, it achieves average relative gains of \textbf{29.73\%} in EM, \textbf{20.82\%} in ES, and \textbf{49.34\%} in BLEU compared to the strongest baseline.
Ablation studies confirm that each component of \approach contributed to the final effectiveness.
Moreover, targeted analyses on specific code structures, as well as experiments across repositories from diverse domains and under varied contextual environments, demonstrate that \approach consistently yields significant gains—highlighting its robustness and generalizability for repository-level code generation.

The main contributions of this paper are as follows:

\begin{itemize}
    \item We propose a novel repository-level code generation framework that situates the target function within its  \textit{upstream} (callers) and \textit{downstream} (callee) contexts.
    \item We are the first to inline a function into its callers' context, allowing the LLM to gain a deeper understanding of the function's intended purpose and usage patterns.
    \item We conduct extensive evaluations on multiple benchmarks. Results demonstrate that our method achieves consistent improvements over state-of-the-art baselines across diverse context environment and across multiple domains.
\end{itemize}

\section{Motivation}

\begin{figure}
    \centering
    \includegraphics[width=1.0\linewidth]{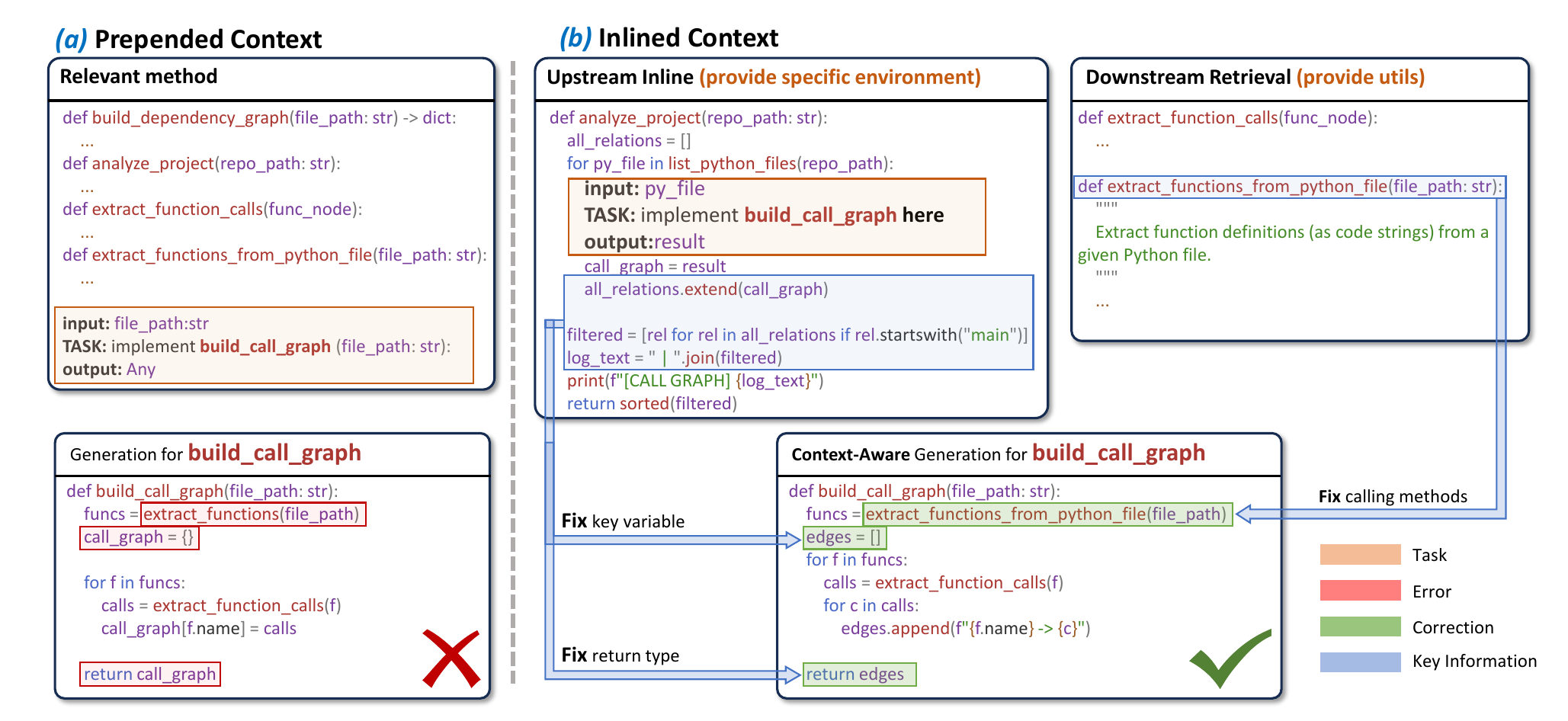}
    \caption{A Motivating Example. Inlining the target function into its call chain creates a more context-aware task formulation, leading to repository-consistent completions.}
    \Description{des}
    \label{fig:motivation_example}
\end{figure}

A central challenge in repository-level code generation lies in effectively incorporating repository context into models~\cite{cheng2024dataflow}. Current approaches typically adopt a retrieval-based framework, where relevant snippets are selected based on text similarity or structural proximity. While this strategy provides local context, it rarely offers a systematic way to represent how a function is actually used by its callers (upstream) or how it depends on its callees (downstream)~\cite{zhang2023repocoder}.

Some recent efforts attempt to incorporate call-chain information, but they often do so by simply prepending retrieved functions to the unfinished target signature~\cite{liu2024graphcoder}. This linear concatenation neglects the target's functional role in its actual calling environment. As a result, the model may remain unaware of input/output constraints or calling conventions, leading to an incomplete or misleading understanding of function semantics. Without adapting retrieved snippets to the actual calling sites, models can frequently fail to infer correct variable bindings, return types, or the appropriate API variants used across the repository~\cite{zhang2024codeagent}.

Figure~\ref{fig:motivation_example} (a) illustrates an example of these shortcomings. Conventional methods often isolate the target function and attach utility functions or snippets above it, failing to capture genuine calling relationships. Consequently, the model incorrectly calls the inappropriate utility \texttt{extract\_functions} and produces a \texttt{dict} instead of the expected \texttt{list[str]}.

This example motivates a new paradigm of context incorporation, as illustrated in Figure~\ref{fig:motivation_example} (b): we can inline the target function directly within its calling context, thereby providing \textit{Inlined Context}. This approach transforms structured call-graph information into a form that the model can readily interpret, exposing crucial signals such as variable bindings, return formats, and valid API usage. As the example shows, the inlined context enables the model to generate repository-consistent code: the correct function now returns a \texttt{list[str]} and invokes the appropriate callees. By coupling targeted context retrieval with inlining, our method enables context-aware, repository-aligned code generation that addresses the shortcomings of conventional retrieval-based strategies.

\section{Methodology}
\label{sec:methodology}

\subsection{Overall Framework}
\label{sec:framework}

Given the signature of an unfinished function \(x\) within a repository $\mathcal{D}$, the goal of repo-level code generation is to produce the function body \(y\) by leveraging contextual information \(\mathcal{C}\) from \(\mathcal{D}\). The main challenge of this task lies in extracting useful context \(\mathcal{C}\) as accurately as possible and enabling the model to comprehend the information within \(\mathcal{C}\). This entails reasoning over entire repositories and understanding complex dependencies across functions, classes, and modules.

\begin{figure}
    \centering
    \includegraphics[width=0.95\linewidth]{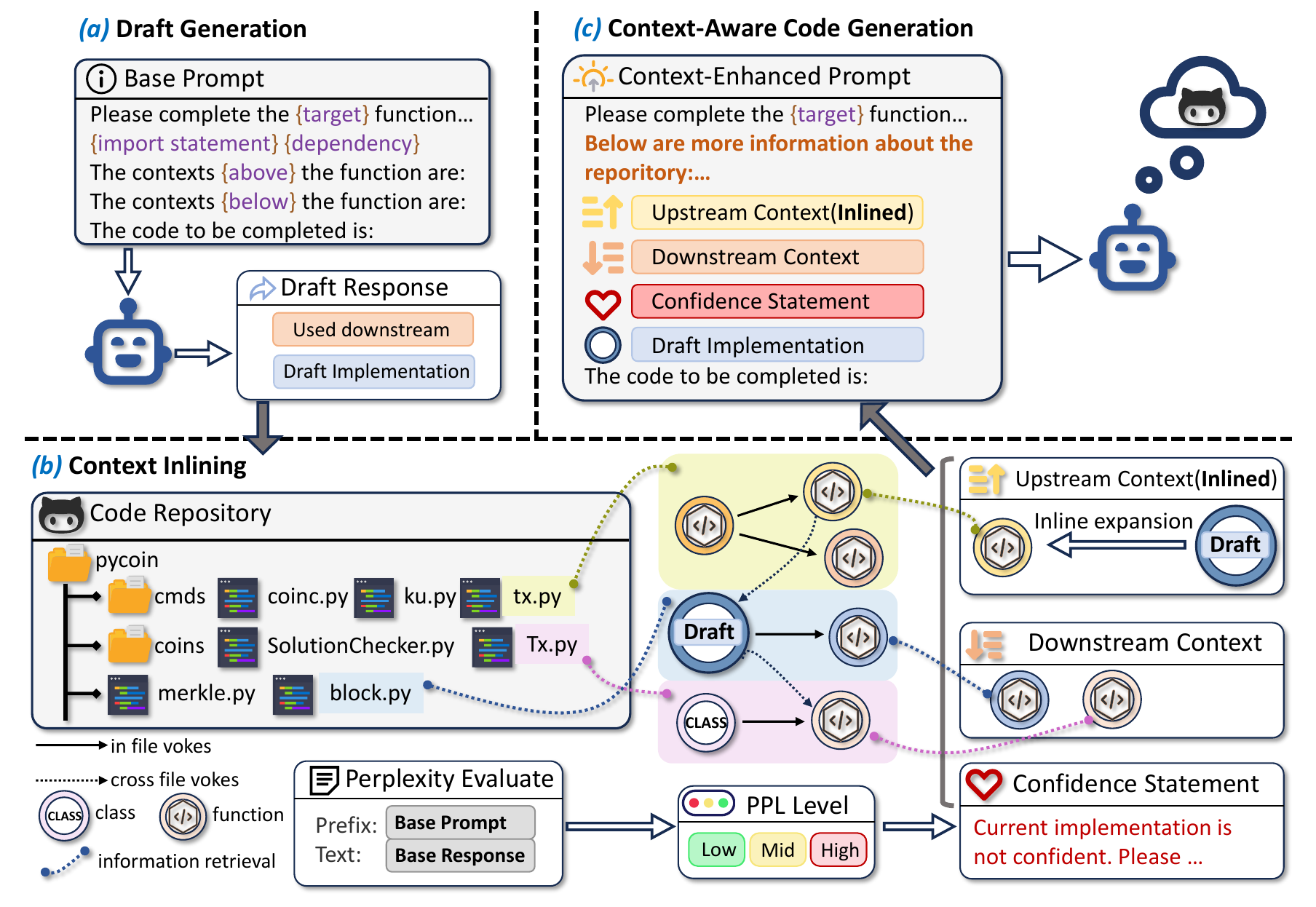}
    \caption{Framework of \approach.}
    \Description{des}
    \label{fig:framework}
\end{figure}

To tackle this challenge, we propose \textit{\approach}, a novel framework for repository-level code generation. Unlike previous techniques that retrieve similar snippets in the context, the core idea of \approach is to situate the target function within its call-graph environment. As illustrated in Figure~\ref{fig:framework}, \approach follows a three-stage pipeline:

1) \upperAnchor Generation (Section \ref{sec:anchor_generation}): Given a base prompt built around the target function signature, \approach first produces a preliminary implementation of the unfinished function, which serves as an anchor. This draft not only offers an approximate view of potential downstream dependencies-helping to situate the target function within the repository's call graph-but also provides a basis for Perplexity evaluation.

2) Context Inlining (Section \ref{sec: retrieval}): The \loweranchor is then inlined into its callers to capture upstream usage scenarios, while relevant callee implementations are retrieved and incorporated downstream. This process yields a coherent, linearized representation that flattens both upstream and downstream contextual information into a unified view for the model.

3) Context Integration (Section \ref{sec: generation}): \approach integrates all relevant information into a context-enhanced prompt that consists of: (a) the \textit{base prompt} (function signature + imports), (b) the retrieved upstream and downstream context, and (c) the \textit{initial \loweranchor}. This enriched prompt provides the LLM with comprehensive guidance, enabling it to generate the final implementation that is both semantically accurate and consistent with the repository environment. The details of each stage are elaborated in the following subsections.

\subsection{\upperAnchor Generation}
\label{sec:anchor_generation}

The \loweranchor generation stage produces a preliminary implementation of the target function, referred to as the anchor. This \textit{\loweranchor} serves as a semantically enriched artifact that guides subsequent retrieval and provides a reference point for the final generation.

To generate this \loweranchor, we first build an initial prompt that offers the LLM a comprehensive view of local context and dependencies. We collect all import statements from the target file and append the full code of all directly referenced dependencies (e.g., variables, functions, classes) discovered within the repository, maintaining the original import order. This cross-file reference information is provided by the datasets (i.e., REPOEXEC~\cite{le2024repoexec} and DevEval~\cite{li2024deveval})
We then add the signature of the unfinished function along with its natural language description.

This structured prompt is then fed to the LLM, which produces both a candidate implementation of the function body (\loweranchor) and a list of the API calls used within it. The resulting \loweranchor plays a dual role: On one hand, it acts as a seed for retrieval; as the identified API calls provide explicit, high-confidence signals for the subsequent \textbf{Downstream Retrieval} process; on the other hand, it serves as a reference for generation, used to calculate a perplexity-based confidence score to inform the final generation stage. It also acts as a \loweranchor implementation that can be refined during final generation.

\subsection{Context Inlining}
\label{sec: retrieval}

Building on the initial implementation during \loweranchor generation stage, we introduce a dual-pronged inlining process to assemble both its upstream and downstream contexts. The core insight is grounded in the observation that a function’s role is defined by its position within the repository’s call graph: its behavior is constrained by its \textit{upstream} callers (how it is used), while its implementation depends on its \textit{downstream} callees (what it depends on).

\subsubsection{Upstream Inlining}
\label{sec:upstream_retrieval}

\begin{figure}[h]
    \centering
    \includegraphics[width=\linewidth]{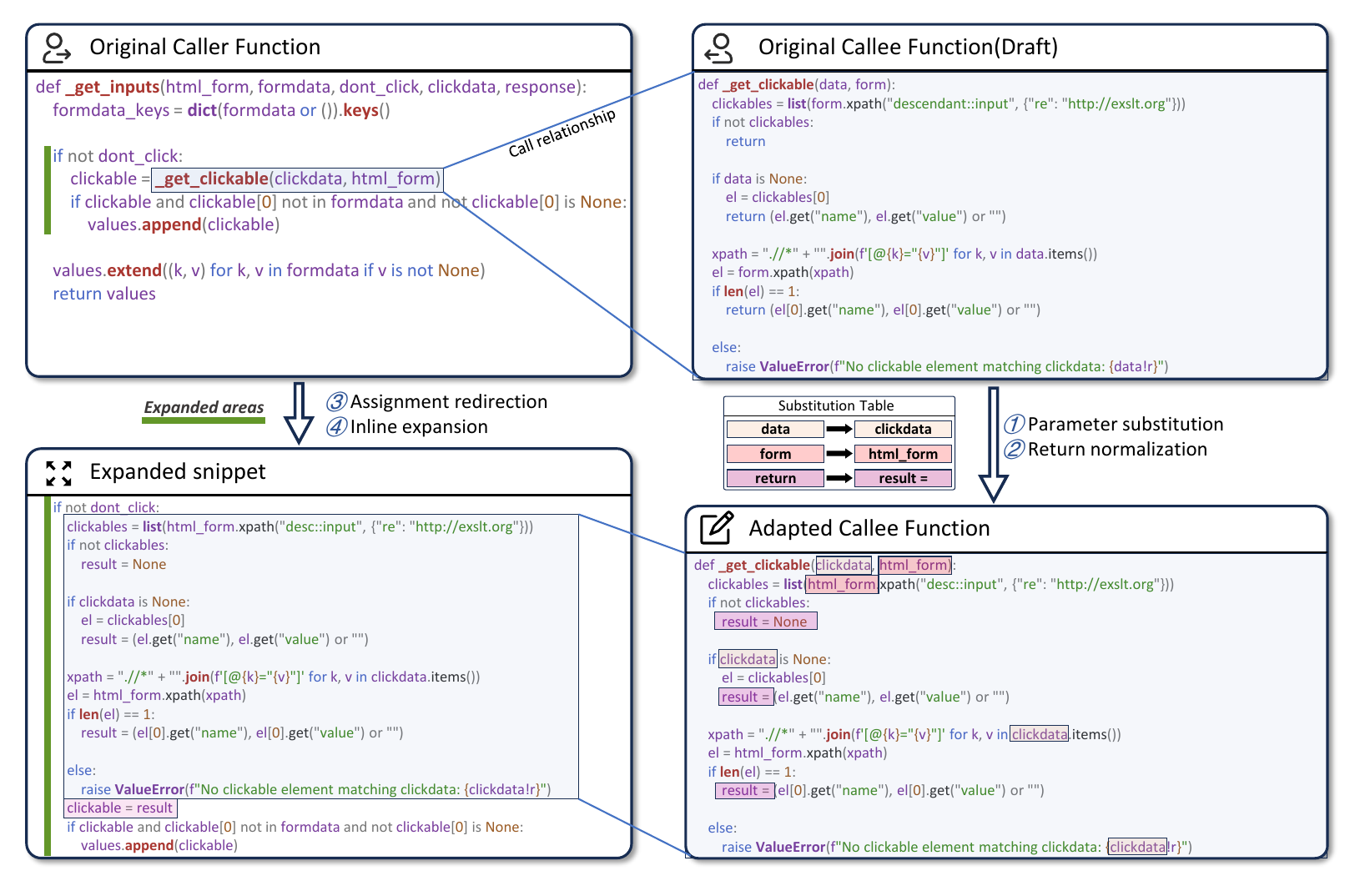} 
    \caption{Function Inlining.}
    \Description{des}
    \label{fig:inline_transformation}
\end{figure}

Upstream inlining aims to provide the LLM with a rich understanding of how the target function is intended to be used. The process starts by identifying all callers of the target function. We traverse the repository's AST to identify all call sites where the target function is invoked. For each call site, we extract the corresponding calling function, referred to as a \textit{caller}.
Unlike traditional methods that merely prepend caller context to the prompt, \approach employs a novel \textbf{context inlining} technique
. As shown in Figure~\ref{fig:inline_transformation}, we embed the \loweranchor directly into its callers, producing a coherent, linearized representation of the function context. This is achieved through a four-step transformation:

\begin{enumerate}

    \item \textbf{Parameter Substitution:}
          Let $\text{Args} = [a_1, a_2, \ldots, a_m]$ denote the argument list at the call site,
          and $\text{Params} = [p_1, p_2, \ldots, p_m]$ denote the formal parameters in the callee definition.
          We define a substitution function over identifiers:
          \begin{equation}
              \label{eq: sigma}
              \sigma : \mathcal{I} \to \mathcal{I}', \quad
              \sigma(x) =
              \begin{cases}
                  a_i & \text{if } x = p_i \in \text{Params}, \\[2mm]
                  x   & \text{otherwise},
              \end{cases}
          \end{equation}
          where $\mathcal{I}$ is the identifier set of the callee body, and $\mathcal{I}'$ denotes the set of identifiers after substitution.

          We lift $\sigma$ to statements, $\mathcal{S} \to \mathcal{S}'$, by recursively applying it to all identifiers in a statement $s \in \mathcal{S}$,
          where $\mathcal{S}$ is the set of statements in the callee body, and $\mathcal{S}'$ denotes the statements after identifier substitution.

          For the callee body $\text{Body}_f = \{s_1, \ldots, s_N\}$, the parameter-substituted body is:
          \begin{equation}
              \sigma(\text{Body}_f) = \{\, \sigma(s) \mid s \in \text{Body}_f \,\} \subseteq \mathcal{S}'.
          \end{equation}

    \item \textbf{Return Normalization:}
          We define a transformation function $\tau: \mathcal{S} \to \mathcal{S}$ operating on statements:
          \begin{equation}
              \label{eq: tau}
              \tau(s) =
              \begin{cases}
                  \texttt{result = exp}  & \text{if } s \equiv \texttt{return $exp$}, \\[4pt]
                  \texttt{result = None} & \text{if } s \equiv \texttt{return},       \\[4pt]
                  s                      & \text{otherwise}.
              \end{cases}
          \end{equation}
          Lifting $\tau$ to statement sets, we obtain the normalized body:
          \begin{equation}
              \tau(\text{Body}) = \{\, \tau(s) \mid s \in \text{Body} \,\}.
          \end{equation}

    \item \textbf{Assignment Redirection:}
          Suppose the original call site has the form
          \texttt{x = f($a_1, \ldots, a_m$)}.
          After parameter substitution and return normalization, this assignment is redirected to
          \texttt{x = result}, binding the call result to the caller’s variable.

    \item \textbf{Inline Expansion:}
          The transformed callee body is obtained by sequentially applying the parameter substitution $\sigma$ (defined in Equation~\ref{eq: sigma}) and return normalization $\tau$  (defined in Equation~\ref{eq: tau}):
          \begin{equation}
              \text{Body}_f^{*} = \tau\big(\sigma(\text{Body}_f)\big).
          \end{equation}
          Inline expansion replaces the original call with $\text{Body}_f^{*}$,
          preserving indentation and surrounding syntactic structure in the caller’s function.

\end{enumerate}

This inlining process ensures semantic equivalence while making the target function's intended behavior explicit within the context. By embedding the draft implementation directly into its upstream caller bodies, the procedure converts a distributed, structured set of cross-file call relations into a single, linearized function-local environment. As a result, the LLM receives a compact, execution-oriented view of the target function's role, facilitating the inference of input/output expectations, preserving the logical flow of computations, and highlighting the relationships between variables and control structures. This transformation also reduces ambiguity and eliminates the distractions inherent in separately retrieved snippets, enabling the model to focus on the essential behavioral patterns of the target function within its repository context.

\subsubsection{Downstream Retrieval}
\label{sec:downstream_retrieval}

\begin{figure}[t]
    \centering
    \includegraphics[width=\linewidth]{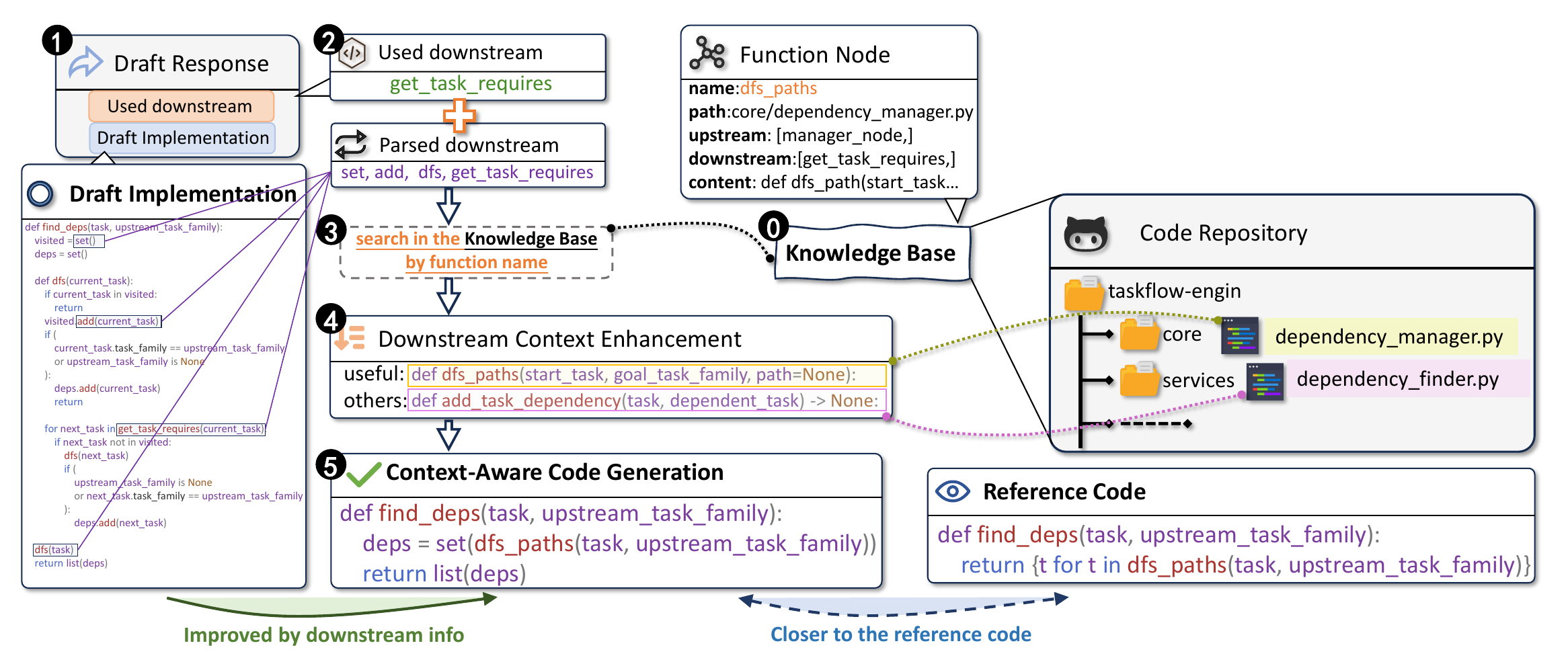} 
    \caption{Downstream Retrieval.}
    \Description{des}
    \label{fig:downstream_workflow}
\end{figure}

Downstream retrieval aims to equip the LLM with the dependent functions of the target function. While the \loweranchor{} may initially implement functionality from scratch, high-quality code often depends on existing functions within the repository. To identify these dependent functions, we consider two complementary sources: (i) parsing the \loweranchor{}'s AST to extract all invoked function calls, denoted as $Q_{\text{AST}}$, and (ii) collecting the LLM-generated list of predicted callees, denoted as $Q_{\text{LLM}}$. We then explicitly take their union to construct the overall query vocabulary:

\begin{equation}
    Q = Q_{\text{AST}} \cup Q_{\text{LLM}} = \{q_1, q_2, \ldots, q_m\},
\end{equation}

where each $q_i$ denotes a candidate function name. Given a repository represented as a collection of function units
$\mathcal{F} = \{f_1, f_2, \ldots, f_n\}$, each with an associated function identifier $\texttt{name}(f_j)$, we apply a substring-based retrieval strategy:
\begin{equation}
    \mathcal{G} = \{ f_j \in \mathcal{F} \;\mid\; \exists q_i \in Q, \; q_i \subseteq \texttt{name}(f_j) \}.\end{equation}
where $\mathcal{G}$ denotes the set of candidate downstream functions. \(\mathcal{G}\) is finally used to form the downstream context, providing enhancements to Prompt.

To prevent data leakage, we exclude the target function itself from $\mathcal{G}$.  The objective is to maximize coverage of potentially useful repository APIs that the target function may leverage, thereby providing the LLM with a richer and more accurate implementation environment.

Thus, downstream retrieval bridges the gap between a noisy, from-scratch draft and a concise, high-quality solution grounded in repository knowledge by ensuring that the retrieved functions are tightly aligned with the model’s actual generation trajectory. This process enables the system to surface the most relevant utility functions exactly when they are needed, avoiding both under-retrieval (missing key APIs) and over-retrieval (introducing irrelevant noise). As a result, the LLM is equipped with a targeted and contextually coherent function set, which maximizes reuse of repository knowledge and ultimately promotes higher-quality, repository-consistent code generation.

\subsection{Context-Aware Code Generation}\label{sec: generation}

Having gathered both the draft implementation and enriched contextual information, we guide the LLM to produce the final code.

We allow the model to operate in a dual mode: it may either reuse the initial draft or refine it based on the enriched context.
The \loweranchor{} serves as a valuable starting point, but it may contain inaccuracies or incomplete logic. To calibrate how much the model should rely on this draft, we introduce a confidence mechanism based on perplexity (PPL)~\cite{jelinek1977perplexity}. For a draft implementation \(R = (r_1, \dots, r_M)\) conditioned on the base prompt \(B\), the PPL is defined as:

\begin{equation}
    \label{eq: ppl}
    \mathrm{PPL}(R \mid B) = \exp\left( -\frac{1}{M} \sum_{j=1}^{M} \log p\bigl(r_j \mid B, r_{<j}\bigr) \right).
\end{equation}
where \(p(r_j \mid B, r_{<j})\) denotes the conditional probabilities estimated by the language model. This formulation provides a direct measure of confidence: lower values of \(\mathrm{PPL}(R\mid B)\) indicate greater model confidence in the draft.

We categorize the model's confidence in the draft code (Section~\ref{sec: generation}) into three levels based on their perplexity: low confidence (PPL\( > \)2), medium confidence (PPL\( \in\)[1.3, 2]), and high confidence (PPL\(< \)1.3). The thresholds are carefully tuned so that roughly 40\% of the samples fall into the low-confidence group, 40\% into the medium-confidence group, and 20\% into the high-confidence group.

For high confidence (low PPL), the model is prompted to trust and closely follow the \loweranchor. When the draft obtains medium confidence, the model is advised to modify or improve the \loweranchor. In cases of low confidence, the model is prompted to critically reassess the \loweranchor, with freedom to regenerate from scratch. Each confidence level is mapped to a natural-language guidance integrated into the final prompt:

\begin{itemize}
    \item \textbf{High confidence}: \textit{``The current implementation and the comments are good, please refer to it and keep these comments.''}
    \item \textbf{Medium confidence}: \textit{``The current implementation is somewhat uncertain and comments are reasonable. Please refer to it partially.''}
    \item \textbf{Low confidence}: \textit{``The current implementation is not confidently correct. Please consider regenerating it.''}
\end{itemize}

\begin{figure}
    \centering
    \includegraphics[width=1.0\linewidth]{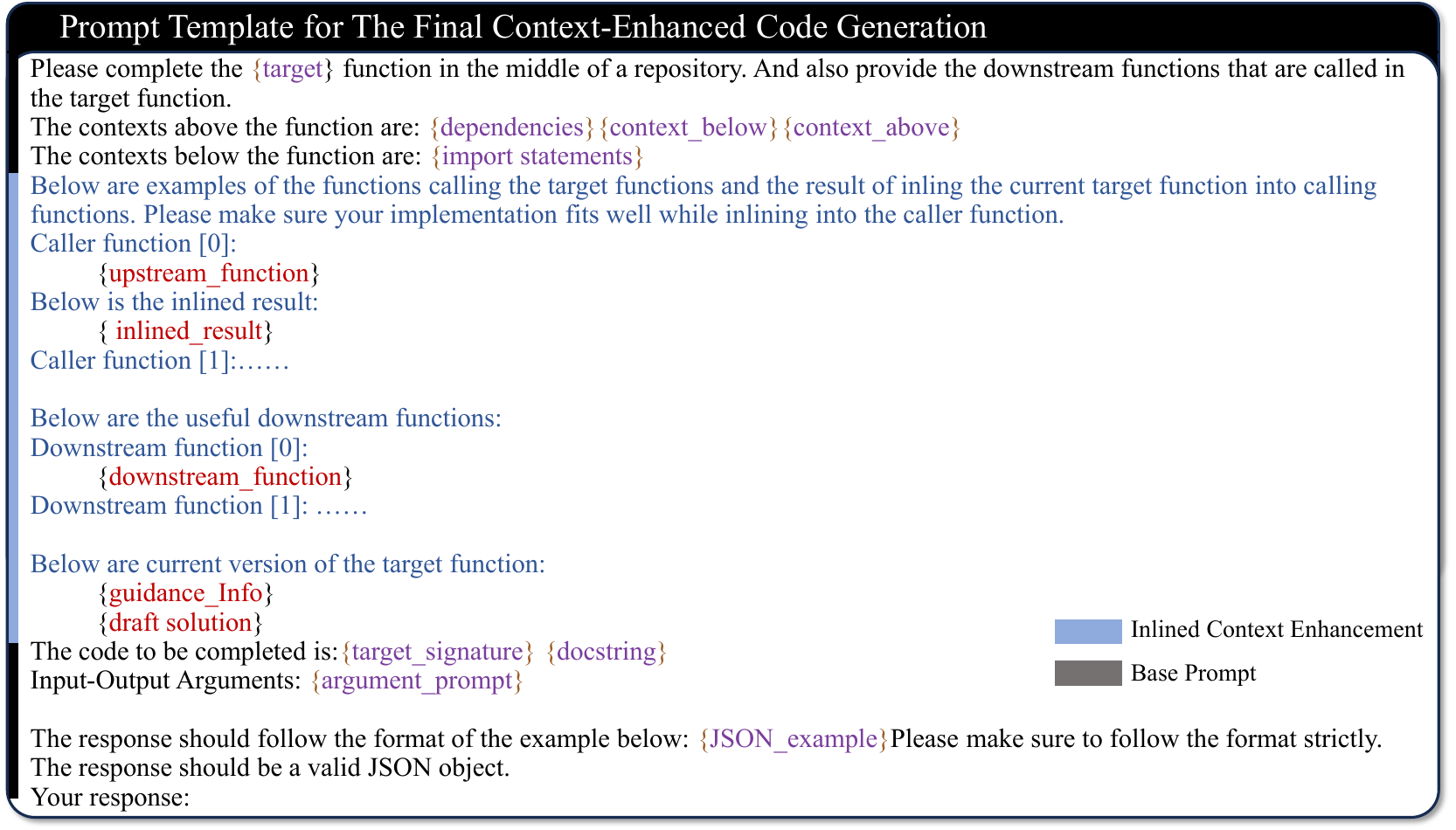}
    \caption{Prompt Template for The Final Context-Enhanced Code Generation.}
    \Description{des}
    \label{fig:prompt_long}
\end{figure}

The final prompt, as illustrated in Figure~\ref{fig:prompt_long}, aggregates all contextual signals, including imports, enriched context, the generation mode guidance, the draft, and the target signature. This structured prompt provides the LLM with a multi-perspective view: repository dependencies, usage patterns, prior draft guidance, and explicit task specifications, enabling the model to generate a final function body that is contextually grounded, semantically consistent, and repository-aware.

Figure~\ref{fig:downstream_workflow} shows a working example illustrating the entire workflow.
(1) The model produces a complex draft that attempts to reimplement a depth-first search (DFS) procedure from scratch, without leveraging existing utilities in the repository.
(2) Based on this draft response, two sources of potential callees are extracted---function calls parsed from the AST and the LLM-predicted callees---which are unified into a query set \(Q\).
(3) This query is then used to search within the repository's knowledge base via substring matching, yielding a candidate set $\mathcal{C}$ of downstream functions. Among these retrieved functions, the system identifies the existing utility \texttt{dfs\_paths}.
(4) In the next stage, this downstream information $\mathcal{C}$ is injected into the prompt as additional context.
(5) The model is then guided to perform the final Context-Aware Code Generation.

\section{Experiment Setup}
\label{sec:experiments}

We comprehensively evaluate the effectiveness of \approach by addressing the following research questions (RQs):
\begin{itemize}
    \item[\textbf{RQ1}] \textbf{(Overall Effectiveness)}: How does \approach perform on repository-level code generation tasks compared to state-of-the-art baselines?
    \item[\textbf{RQ2}] \textbf{(Ablation Study)}: To what extent do the key components of \approach contribute to its overall performance?
    \item[\textbf{RQ3}] \textbf{(Qualitative Analysis)}: How does the upstream-downstream retrieval mechanism in \approach provide effective context for code generation?
    \item[\textbf{RQ4}] \textbf{(Domain Generalization)}: How does the performance of \approach generalize across different programming domains?

\end{itemize}

\subsection{Datasets}
\label{sec:datasets}

We conduct our experiments on two prominent repository-level code generation benchmarks: DevEval~\cite{li2024deveval} and REPOEXEC~\cite{le2024repoexec}. We focus on completing the entire body of the unfinished functions.

\begin{itemize}
    \item{\textbf{DevEval}} is a repository-level benchmark that aligns with real-world codebases, providing 1,825 annotated samples in Python from 115 repositories across 10 domains to systematically evaluate LLMs’ coding abilities.
    \item{\textbf{REPOEXEC}} is a benchmark for repository-level code generation that provides 355 functions in Python with ground-truth annotations.
\end{itemize}

\subsection{Evaluation Metrics}
\label{sec: metrics}

Following established practices in repository-level code generation research~\cite{zhang2023repocoder, liu2024graphcoder, cheng2024dataflow}, we evaluate the quality of the generated code from multiple dimensions using the following metrics. Let $y$ denote the reference code snippet (ground truth), and $\tilde{y}$ denote the generated code produced by the model. Identifiers such as variable names, API calls, and function names are extracted from a code snippet $y$ into a set $I(y)$. These definitions form the basis for the following evaluation metrics.

\begin{itemize}
    \item \textbf{Exact Match (EM)} checks whether the generated code exactly matches the reference snippet, yielding a binary score.

    \item \textbf{Edit Similarity (ES)}~\cite{lcvenshtcin1966binary} provides a finer-grained measure based on Levenshtein distance, computed as
          \begin{equation}
              \mathrm{ES}(y,\tilde{y}) = 1 - \frac{\mathrm{Lev}(y,\tilde{y})}{\max(|y|,|\tilde{y}|)},
          \end{equation}
          where $\mathrm{Lev}(y,\tilde{y})$ denotes the Levenshtein distance between $y$ and $\tilde{y}$, defined as the minimum number of insertions, deletions, or substitutions required to transform $y$ into $\tilde{y}$.

    \item \textbf{BLEU}~\cite{papineni2002bleu} evaluates the n-gram overlap between the candidate and the reference, defined as
          \begin{equation}
              \mathrm{BLEU} = \mathrm{BP} \cdot \exp \left( \sum_{n=1}^{N} w_n \log p_n \right),
          \end{equation}
          where $p_n$ is the modified n-gram precision, $w_n$ are weights (typically uniform, $w_n = \tfrac{1}{N}$), and $\mathrm{BP} = \min\left(1, e^{1 - \tfrac{|\tilde{y}|}{|y|}}\right)$ is the brevity penalty.

    \item \textbf{Identifier Match F1 (ID.F1)}~\cite{liu2024graphcoder} evaluates identifier-level overlap, such as API and variable names, with the F1 score. Let $I(y)$ and $I(\tilde{y})$ denote the sets of identifiers extracted from the reference and generated code, respectively. Precision, recall, and the ID.F1 score are defined as
          \begin{equation}
              Precision = \frac{|I(y) \cap I(\tilde{y})|}{|I(\tilde{y})|}, \quad
              Recall = \frac{|I(y) \cap I(\tilde{y})|}{|I(y)|}, \quad
              \mathrm{ID.F1} = \frac{2 \cdot Precision \cdot Recall}{Precision + Recall}.
          \end{equation}

\end{itemize}

We did not adopt execution-based metrics such as Pass@k (the passing rate of generated code on test cases). Unlike programming-contest benchmarks~\cite{nijkamp2022codegen, du2024evaluating}, repository-level code completion tasks differ substantially in structure and dependencies~\cite{gu2025effectiveness}. In our experiments, we also observed that execution-based scores fluctuate considerably across runs. The same model outputs often result in test failures due to subtle environmental issues that are unrelated to the model itself. These fluctuations could lead to misleading or non-comparable Pass@k results.

\subsection{Baselines}
\label{sec:baselines}

We compare \approach against a suite of baselines spanning a large variety of techniques, including in-file context methods, text-similarity-based retrieval, and static-analysis-based approaches.
\begin{itemize}
    \item{\textbf{In-File}} leverages the full in-file context without utilizing any cross-file information.
    \item{\textbf{Vanilla}} follows the basic prompt construction provided in existing benchmarks such as DevEval~\cite{li2024deveval} and RepoExec~\cite{le2024repoexec}, where related repository context is merely prepended to the target function without deep integration.
    \item{\textbf{RepoCoder}~\cite{zhang2023repocoder}} is a retrieval-augmented framework that iteratively combines similarity-based retrieval with code LLMs to exploit repository-level context for code completion. Specifically, we implemented its function body completion pipeline, using UniXcoder\footnote{\url{https://huggingface.co/microsoft/unixcoder-base}}~\cite{guo2022unixcoder} as the embedding model. For evaluation, we adopted the results from the third retrieval-iteration stage, as reported to be the most effective in the original work.
    \item{\textbf{DRACO}~\cite{cheng2024dataflow}} enhances repository-level code completion via dataflow-guided retrieval, constructing a repo-specific context graph to provide precise and relevant information for LLMs. We adapted the prompt template of this baseline method to support the complete function body completion task.
    \item{\textbf{GraphCoder}~\cite{liu2024graphcoder}} employs a graph-based retrieval-generation framework using code context graphs and a coarse-to-fine retrieval process to acquire repository-specific knowledge more effectively. We have adapted the prompt template of this baseline method to support the complete function body completion task.

\end{itemize}

\subsection{Implementation Details}
\label{sec:implementation_details}

For the \loweranchor generation stage, we employed the identical Vanilla prompts used in the baseline methods for each dataset: the basic prompt construction methods provided by DevEval and REPOEXEC, respectively.

For context retrieval, we parsed Python code and built call graphs with the assistance of Tree-Sitter\footnote{ \url{https://github.com/tree-sitter/tree-sitter}} and Pydepcall\footnote{ \url{https://github.com/FSoft-AI4Code/pydep}}.

To evaluate our framework across different models, we used three advanced backbone LLMs: \texttt{DeepSeek-V3}\footnote{\url{https://huggingface.co/deepseek-ai/DeepSeek-V3-0324}}~\cite{liu2024deepseek}, \texttt{Qwen3-Coder}\footnote{\url{https://huggingface.co/Qwen/Qwen3-Coder-480B-A35B-Instruct}}~\cite{yang2025qwen3}, and \texttt{GPT-5-mini}\footnote{gpt-5-mini-2025-08-07 from~\url{https://platform.openai.com/docs/models/gpt-5-mini}}~\cite{openai2025introducing}. We employed Qwen2.5-Coder-1.5B\footnote{\url{https://huggingface.co/Qwen/Qwen2.5-Coder-1.5B}}~\cite{hui2024qwen2} as the probability estimator for confidence estimation in the final stage. We configured each model to its native context window size to accommodate long input sequences, as our evaluation tasks require substantial context but generate concise outputs. The configured input limits are \SI{128000}{tokens} for \texttt{DeepSeek-V3}, \SI{262144}{tokens} for \texttt{Qwen3-Coder}, and \SI{400000}{tokens} for \texttt{GPT-5-mini}. Output generation was limited to \SI{10000}{tokens} for all models to ensure focused completions. To ensure deterministic and reproducible decoding, all models used a temperature of 0.0, rendering the \textit{top-$p$} sampling parameter ineffective. All experiments were run on a compute server with Intel Xeon Silver 4214R CPUs and NVIDIA A40 GPUs.

\section{Experiments Results}

\subsection{RQ1: Overall Effectiveness}

Table~\ref{tab:total_dev_eval} and \ref{tab:total_repo_exec} compare the performance of various methods on the DevEval and RepoExec datasets across three backbone models.

\input{asserts/tables/total_dat_dev_eval}
\input{asserts/tables/total_data_repo_exec}

Overall, \approach achieves substantial and stable improvements across all three backbones. Notably, \approach achieves stable improvement compared to Vanilla, a baseline method which is also used in the draft generation stage of \approach. To quantify the gains, we compute the relative percentage improvements over the strongest baseline for each backbone model, and then report the averaged results across the three models. On DevEval, the average relative improvements reach \textbf{5.13\%} in EM, \textbf{10.86\%} in ES, and \textbf{10.67\%} in BLEU; on RepoExec, the gains are even larger, with \textbf{29.73\%} in EM, \textbf{20.82\%} in ES, and \textbf{49.34\%} in BLEU, highlighting \approach's robustness across models and datasets.

The In-File baseline—using only intra-file context—remains competitive, sometimes outperforming GraphCoder and DRACO. This suggests that preserving local context is critical. By contrast, GraphCoder and DRACO often underperform because they were originally designed for line-level completion and rely on assumptions (e.g., partial function bodies or dataflow analysis) that do not hold in function-body generation. Additionally, they discard in-file context, which can contain crucial local information, further limiting their effectiveness in repository-level code generation.

We notice that the EM scores in our experiments are relatively lower compared to those reported in prior work. This discrepancy arises from differences in completion objectives: while previous studies primarily evaluate single-line completion, our task requires completing entire function bodies, which is inherently more challenging.

\answerbox{Answer to RQ1}{
    InlineCoder substantially outperforms all baseline methods, demonstrating consistent improvements across diverse datasets and models. The results highlight the effectiveness of context inlining in improving repository-level code generation.
}

\subsection{RQ2: Ablation Study}
\label{sec:ablation}

\input{asserts/tables/ablation_dev_eval}
We conduct an ablation study on the \textsc{DevEval} dataset using the DeepSeek-V3 backbone. The ablated variants include: removing \emph{upstream context retrieval} (\textsc{w/o upstream}), removing \emph{usage-context inlining} (\textsc{w/o inline}), removing \emph{downstream context retrieval} (\textsc{w/o downstream}), removing the \emph{confidence statement} derived from the \loweranchor's perplexity (\textsc{w/o confidence}), and removing the \emph{\loweranchor implementation} itself from the context-aware code generation (\textsc{w/o draft}).

As shown in Table~\ref{tab:ablation_DevEval}, eliminating any key component results in performance degradation, which confirms their effectiveness. Removing the  \emph{\loweranchor implementation} causes the largest drop, showing that anchoring generation to concrete candidates is crucial for accuracy and semantic alignment. The removal of \emph{inlining} and \emph{upstream/downstream context} also consistently degrades performance, underscoring their complementary roles. Notably, the performance of \textsc{w/o inline} is comparable to that of \textsc{w/o upstream}, suggesting that upstream information alone provides limited benefit if not properly linearized. Finally, removing the \emph{confidence statement} leads to moderate but stable drops, showing that perplexity-based scoring helps select more effective prompts.

\answerbox{Answer to RQ2}{
    Each proposed component of the \approach contributes positively to performance. The \emph{\loweranchor implementation} plays the most critical role, and \emph{ function usage context inlining} is essential for effectively leveraging context information, and cross-file retrieval (upstream and downstream) provides complementary gains. This confirms that \approach benefits from a carefully designed integration of contextual signals.
}

\subsection{RQ3: Qualitative Analysis}

\input{asserts/tables/validation_upstream}

\subsubsection{Improved performance on return statements}
\label{sec:upstream_eval}

To further investigate the role of upstream context in function completion, we focus on the generation of the \texttt{return} statement, which is typically located at the last line of the target function. Specifically, we extract the final line of code from both the baseline generations and \approach, and evaluate their accuracy against the reference code. Following common practice in single-line code evaluation, we adopt EM, BLEU, and ES as metrics, all of which were introduced in Section~\ref{sec: metrics}. Experiments are conducted on the \textsc{DevEval} dataset using the DeepSeek-V3 backbone. We compare \emph{\approach} with two representative methods: Vanilla and RepoCoder (a strong baseline leveraging multiple generations).

As shown in Table~\ref{tab:validation_upstream}, \approach consistently achieves the best performance across all metrics.
In particular, \approach improves EM scores by 2.07\% on return statements.
While RepoCoder also leverages iterative generation, it still falls short of InlineCoder, suggesting that naive multi-generation alone is insufficient. Instead, incorporating structured upstream usage information provides more reliable guidance for producing accurate final statements. These findings confirm that upstream information plays a crucial role in shaping the correctness and naturalness of the return expressions, complementing the benefits of inlining and retrieval strategies demonstrated in earlier experiments.

\input{asserts/tables/validation_downstream}

\subsubsection{Improved performance on function-call statements}
\label{sec:downstream_eval}

To investigate the role of downstream information in improving function-call correctness, we conducted a targeted evaluation on the generated invocation statements using the \textsc{DevEval} dataset with the DeepSeek-V3 backbone. For each generated function, we extracted all invocation statements and compared them against the reference annotations.
We adopted several evaluation metrics to assess this alignment, including EM, Jaccard Similarity, F1 Score, Coverage, and Downstream Invocation Recall (DIR)~\cite{le2024repoexec}. These metrics collectively measure the accuracy and completeness of both callee names and full invocation instances.
For comparison, we also conducted experiments on Vanilla, RepoCoder, and \approach.

As shown in Table~\ref{tab:validation_downstream}, the proposed context-aware generation (\emph{InlineCoder}) achieves the overall best performance. In particular, \approach improves EM by 4.27\% on function-call statements.
These results demonstrate that incorporating downstream information substantially improves the correctness and coverage of invocation statements, thereby strengthening the functional reliability of generated code.

\answerbox{Answer to RQ3}{

    InlineCoder significantly improves the accuracy of both return and function-call statements across multiple evaluation metrics, demonstrating the effectiveness of its novel approach to leveraging upstream and downstream information through inlined context.
}

\subsection{RQ4: Domain Generalization}

\begin{figure}
    \centering
    \includegraphics[width=1.0\linewidth]{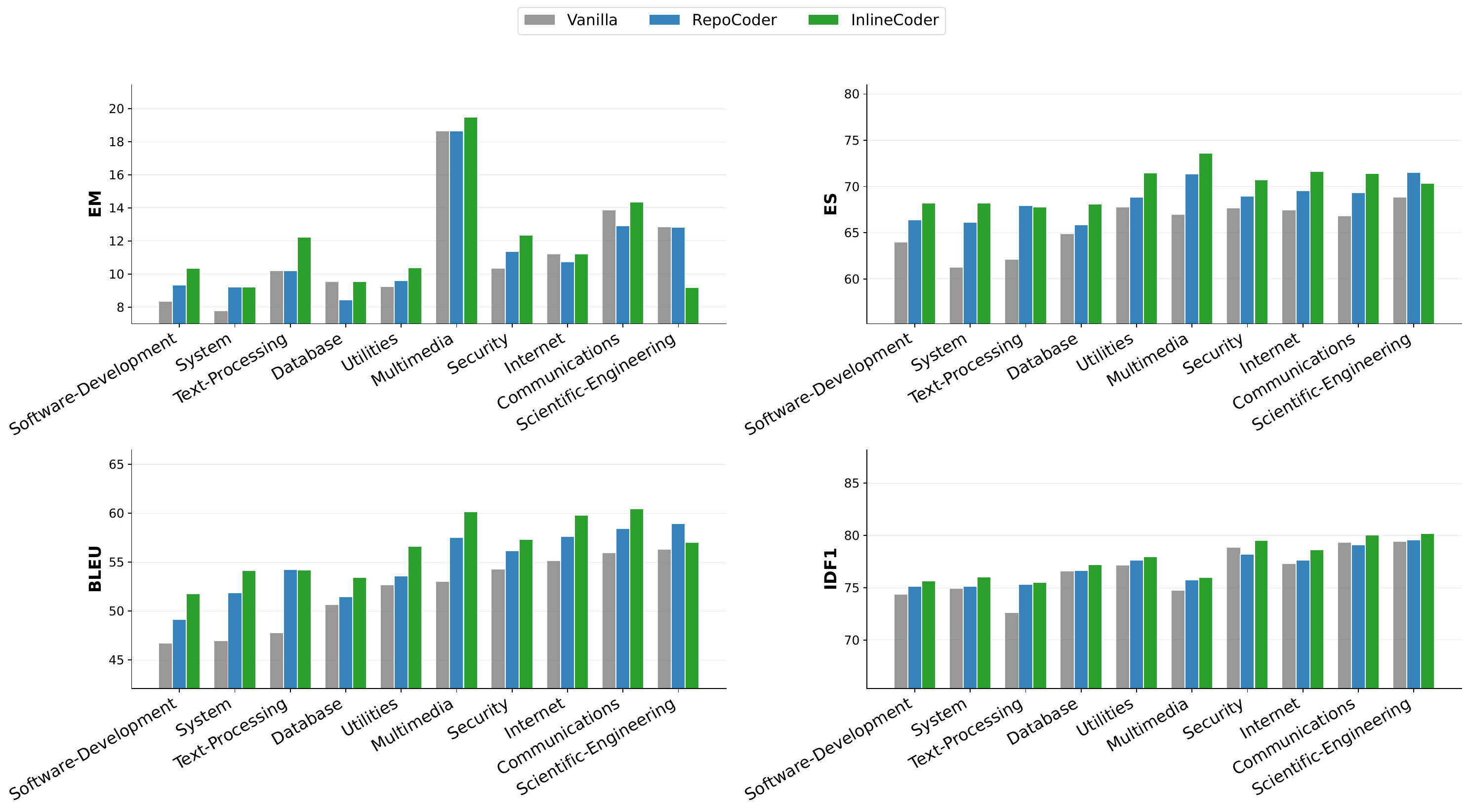}
    \caption{Comparison of Effectiveness across Various Domains.}
    \Description{des}
    \label{fig:domain_deveval}
\end{figure}

To further investigate the generalizability of our approach across different domains, we adopt the domain taxonomy provided by the \textsc{DevEval}~\cite{li2024deveval} dataset and evaluate four metrics: EM, ES, BLEU, and ID.F1. Experiments are conducted on the DevEval dataset using the DeepSeek-V3 backbone. The results over ten domains are summarized in Figure~\ref{fig:domain_deveval}.

In terms of Exact Match (EM), \approach achieves the best performance in 9 out of 10 domains. For ES, it outperforms the baselines in 8 domains. On BLEU, \approach leads in 9 domains, while on ID.F1, it achieves the best results across all 10 domains. Compared to the Vanilla baseline, \approach demonstrates consistent improvements in almost all domains.

The only relatively weaker performance is observed in the \textit{Scientific-Engineering} domain, which may be attributed to the nature of this domain: its tasks are primarily oriented towards scientific computing, where the cross-function invocation structure is relatively sparse. Consequently, the downstream and upstream information leveraged by our method provides limited additional benefit in this domain.

\answerbox{Answer to RQ4}{
    InlineCoder demonstrates robust generalization across diverse domains, consistently surpassing baselines in nearly all settings. The results validate that inlined upstream and downstream signals provide reliable improvements even under varied repository structures.
}

\subsection{Case Study}

\begin{figure}
    \centering
    \includegraphics[width=1.0\linewidth]{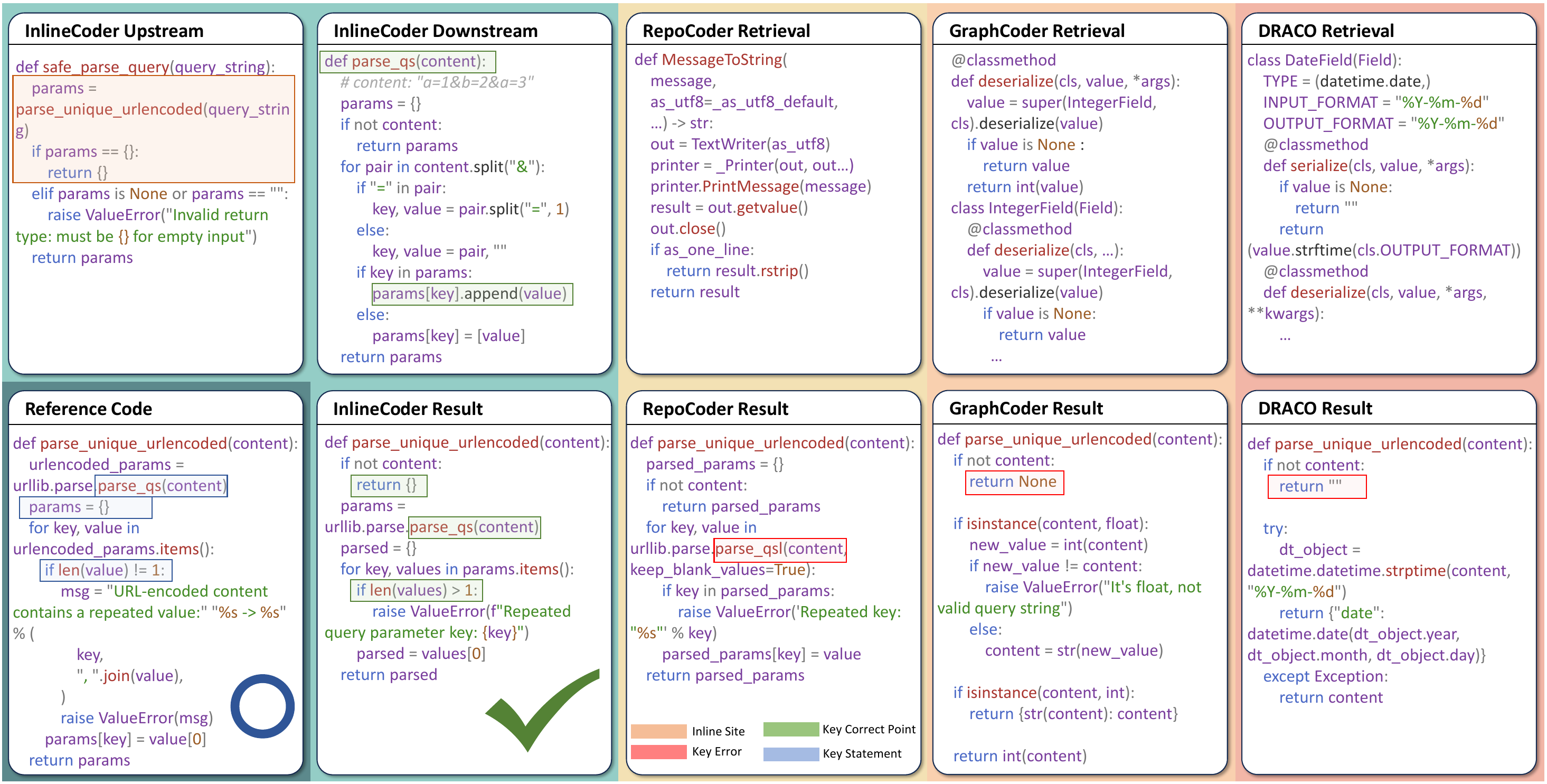}
    \caption{A case about the effectiveness of the bidirectional call inlining.}
    \Description{des}
    \label{fig: case_study}
\end{figure}

Figure~\ref{fig: case_study} presents a case study comparing \approach against other baselines. The target function for completion is \texttt{parse\_unique\_urlencoder}. The figure is organized in a 2×5 grid, where the top row displays key contextual information retrieved by different approaches from the repository, while the bottom row shows the ground truth reference implementation alongside generations from \approach and baseline methods.

This case study demonstrates the effectiveness of \approach's in capturing both upstream and downstream contexts. Through function inlining, \approach successfully identifies and leverages crucial contextual information that directly impacts implementation correctness.

In contrast, baseline approaches such as RepoCoder, which relies on similarity-based retrieval, and DRACO, which employs dataflow analysis, retrieve contextual information that, although structurally relevant, offers limited utility for the specific implementation task.

The advantages of our approach are particularly evident in two key aspects. First, in handling downstream dependencies, only \approach correctly utilizes the \texttt{parse\_qs} utility function with proper parameters. While RepoCoder attempts to use a similarly named \texttt{parse\_qsl} function, its parameter usage differs significantly from the reference implementation. This precision in API usage stems directly from \approach's downstream retrieval mechanism, which not only identifies the correct utility function but also captures its implementation details, ensuring accurate function names and parameter specifications.
Second, \approach demonstrates effectiveness in handling return value. As the case shows, \approach consistently generates the correct return type \texttt{dict{}}. In contrast, baseline methods exhibit inconsistent behavior—sometimes returning \texttt{dict{}}, but also frequently producing \texttt{str}, \texttt{int}, or \texttt{None}. This accuracy in return type handling is achieved through \approach's upstream inlining, which identifies the caller function \texttt{safe\_parse\_query}.

\begin{figure}
    \centering
    \includegraphics[width=1.0\linewidth]{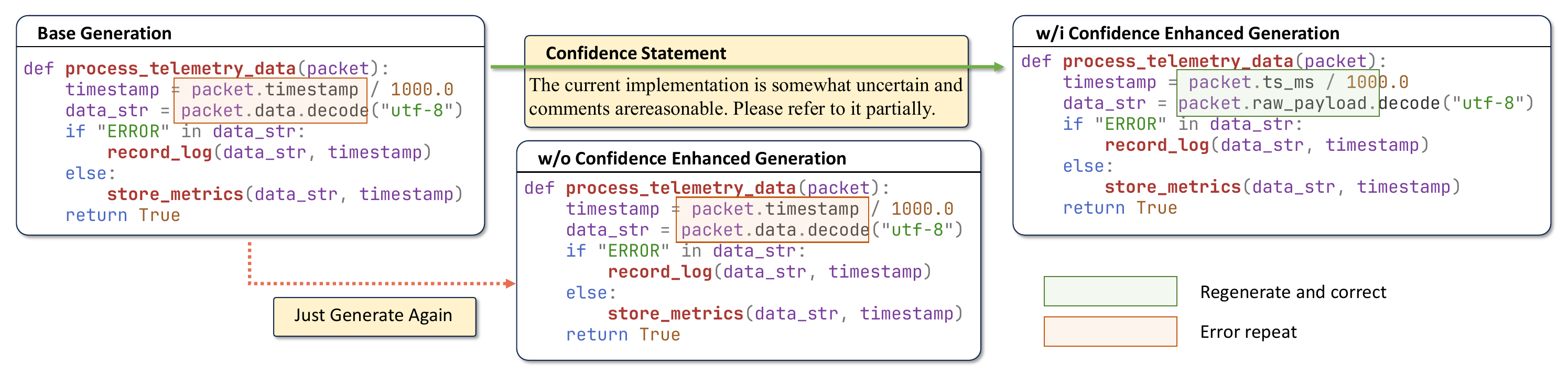}
    \caption{A case about the impact of confidence guidance on mitigating self-repetition bias.}
    \label{fig:confidence_case}
\end{figure}

Figure~\ref{fig:confidence_case} provides another case about how our confidence statement mitigates anchoring bias. In ``Base Generation,'' the LLM defaults to general priors (e.g., \texttt{packet.timestamp}) instead of capturing enhanced context information. Without confidence guidance, the model exhibits strong self-repetition, replicating its initial erroneous draft verbatim despite the presence of the correct inlined context.

Conversely, when \approach identifies the draft as a medium-confidence result, it prepends a confidence statement. This re-frames the task from simple generation to active modification, forcing the LLM to re-evaluate its draft against the inlined information. Consequently, the model successfully identifies structural discrepancies and adopts correct conventions, demonstrating that while context inlining provides the necessary knowledge, confidence guidance serves as the essential trigger to ensure repository-level information is utilized rather than overshadowed by initial model biases.

\section{Discussion}

\subsection{Performance Analysis Across Various Context Environments}

\begin{figure}
    \centering
    \includegraphics[width=1.0\linewidth]{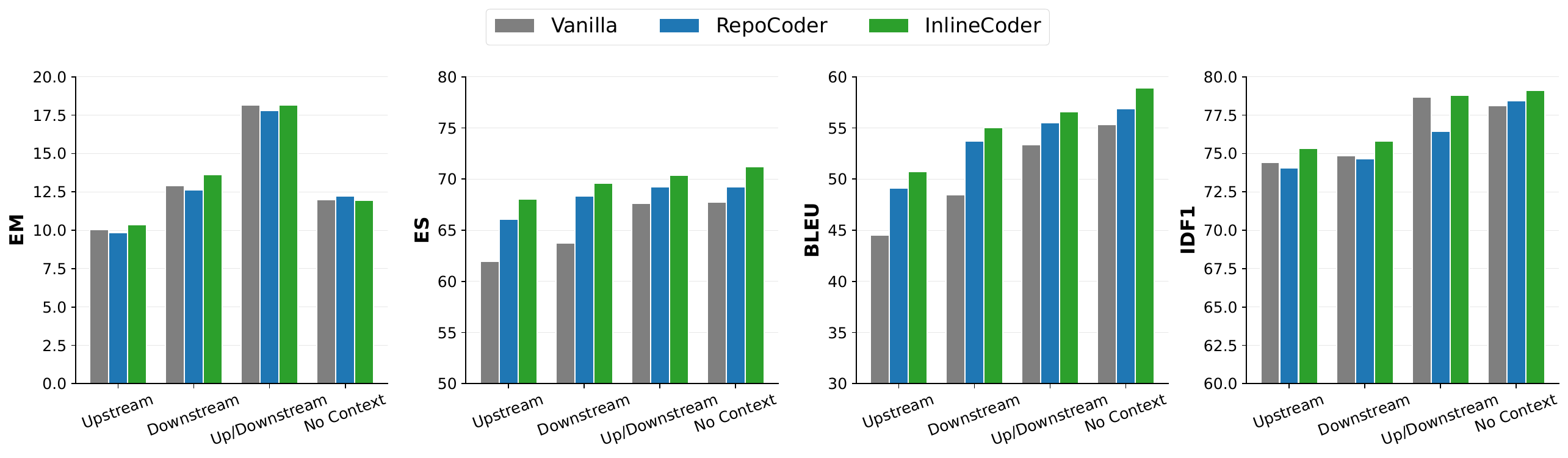}
    \caption{Effectiveness Comparison in Different Context Environments.}
    \Description{des}
    \label{fig:grouped_metrics}
\end{figure}

Figure~\ref{fig:grouped_metrics} illustrates how repository environments influence the effectiveness of \approach. To quantify this impact, we performed a stratified analysis on the \textsc{DevEval} dataset --- a benchmark curated from real-world Python projects on GitHub --- using DeepSeek-V3, categorizing samples into four groups based on their inherent structural characteristics: \textit{Upstream} (16.38\%), \textit{Downstream} (15.29\%), \textit{Up/Downstream} (6.03\%), and \textit{No Context} (62.30\%). Here, the \textit{No Context} group represents isolated functions that naturally lack caller-callee dependencies in their original repositories, rather than intentionally having their contexts removed.

The results show that \approach consistently outperforms baselines in nearly all scenarios. Notably, we observe that overall scores in the \textit{No Context} setting are generally higher than those in context-dependent categories. This is attributed to the fact that functions without upstream or downstream relations are typically independent, resulting in a lower reliance on repository-wide information. The inherent simplicity of these standalone tasks leads to higher baseline performance. Conversely, the lower scores observed in complex repository environments (i.e., those with caller-callee relationships) suggest that the primary challenge lies in the model's difficulty in comprehending and integrating intricate structural contexts.

Furthermore, across the four categories, \approach achieves the most significant performance gain in the \textit{Upstream} setting. This substantial improvement empirically validates the effectiveness of our design in capturing and utilizing upstream caller information.

\subsection{Threats to Validity}
\label{sec:threats}

Like most existing repository-level code generation benchmarks~\cite{jimenez2023swe,ding2023crosscodeeval,li2024evocodebench,li2024deveval,le2024repoexec,yu2024codereval,zhang2023repocoder}, our empirical evaluation was conducted on Python repositories, which may limit the generalizability of our findings to other programming languages or different types of software projects.

We mitigate this threat in two ways. First, we evaluate our method on two distinct datasets that span multiple domains and project scales. Second, the fundamental principle of \approach-leveraging upstream usage context and downstream dependency context-is language-agnostic. The framework relies on Abstract Syntax Tree (AST) analysis, a standard technique applicable to virtually all modern programming languages. Although our implementation is specific to Python, the core methodology can be readily extended to other programming languages such as Java, C++, or TypeScript. Furthermore in strongly typed languages such as C++, AST analysis can provide even more precise and downstream information.

\section{Related Work}
\label{sec:related}

Code Large Language Models (LLMs), such as Code Llama~\cite{roziere2023code}, DeepSeek-Coder~\cite{guo2024deepseek}, and Qwen-Coder~\cite{hui2024qwen2}, have demonstrated remarkable potential in automating software development tasks~\cite{barone2017parallel,becker2023programming,gee2024code, huang2023enhancing,luo2023wizardcoder,zheng2024opencodeinterpreter,zhong2024debug,shi2024between,olgaard2009automated,olgaard2010optimizations,pujar2023automated,vadisetty2023leveraging,chen2025swe,li2025swe}. By internalizing extensive code knowledge from vast corpora into billions of parameters, these models can solve a wide array of general-purpose programming problems~\cite{kazemitabaar2023studying, shi2024code, zeng2025pruning, liventsev2023fully, shi2023large}. Building upon these foundational capabilities, repository-level code generation has evolved significantly~\cite{zhang2023repocoder, shrivastava2023repofusion, li2024deveval, le2024repoexec}. Prior work in this domain can be broadly categorized into retrieval-augmented generation~\cite{zhang2023repocoder, shrivastava2023repofusion}, graph-based and structured retrieval~\cite{liu2024graphcoder, codexgraph, bi2024iterative}, agentic/iterative refinement~\cite{codeRAG, deshpande2024class, bairi2024codeplan}, static-analysis-informed prompting~\cite{liu2024stall+, cheng2024dataflow}, and fine-tuning~\cite{rtlrepocoder, wang2024rlcoder}.

\paragraph{Retrieval-augmented generation.}
Early efforts demonstrated that retrieving relevant code snippets from the repository can substantially improve generation quality. Methods such as RepoCoder~\cite{zhang2023repocoder}, RepoFuse~\cite{liang2024repofuse}, and RepoFusion~\cite{shrivastava2023repofusion} leverage retrieval mechanisms to supply the model with repository-wide context. LongCodeZip~\cite{shi2025longcodezip} selects multiple relevant contexts based on the inherent perplexity from LLMs. Other works explore prompt selection strategies to choose the most useful snippets from large repositories~\cite{shrivastava2023repository}. These approaches establish the value of repository-level context but typically rely on similarity or topical relevance rather than relational call-graph signals.

\paragraph{Graph-based and structured retrieval.}
To capture structural relationships beyond lexical similarity, a line of work constructs graph representations of code. CodexGraph~\cite{codexgraph} builds code graph databases that enable structured querying, while GraphCoder~\cite{liu2024graphcoder} models control-flow and data dependencies using context graphs. RepoHyper and related approaches use semantic repo-level graphs with search-expand-refine strategies to locate relevant code elements~\cite{bi2024iterative}. These methods enhance retrieval precision by leveraging structural relationships; however, they are primarily oriented toward identifying structurally similar or semantically related code snippets, rather than explicitly incorporating upstream or downstream usage signals into the prompts.

\paragraph{Agentic and iterative refinement.}
Agentic frameworks and iterative planners perform multi-step reasoning and use external tools (such as static analysis, testing, or execution) to refine outputs. Examples include CodeRAG, which combines dual graphs with agentic reasoning and specialized tools for graph traversal and code testing~\cite{codeRAG}; RRR, which allows iterative exploration using static analysis~\cite{deshpande2024class}; and CodePlan, which frames repo-level coding as a planning problem with incremental dependency analysis~\cite{bairi2024codeplan}. These works highlight the benefit of multi-step problem decomposition but do not systematically leverage caller-callee signals encoded by call graphs as prompt augmentations.

\paragraph{Static analysis and context pruning.}
Several works incorporate static analysis to prune or enrich the prompt context. STALL+ integrates static analysis into the prompting, decoding, and post-processing stages~\cite{liu2024stall+}; DRACO uses dataflow-guided retrieval to focus on flow-relevant fragments~\cite{cheng2024dataflow}; and hierarchical methods model repositories at function granularity with topological dependencies to reduce noisy context~\cite{hierarchicalcontext}. These techniques enhance relevance and reduce noise. Nevertheless, they remain primarily focused on context selection or compression, rather than the injection of explicit upstream/downstream usage information or confidence signals derived from preliminary model outputs.

\paragraph{Fine-tuning methods.}
Other lines of work focus on domain-specialized model training strategies~\cite{phan2025repohyper}. For instance, RTLRepoCoder fine-tunes for Verilog completion~\cite{zhang2023repocoder}, while curriculum datasets are designed to target hard patterns~\cite{sagtani2024improvingfimcodecompletions}. There are also efforts to improve retrievers (e.g., via reinforcement learning) and to combine retrieval with reinforcement or reflexive training~\cite{wang2024rlcoder,repogenreflex}.

In contrast to these works, \approach introduces several key novelties. First, unlike traditional RAG or graph-based methods that rely on surface-level similarity or static structures, our approach reframes repository-level generation as a function-level task by inlining the unfinished function into its call stack, capturing both upstream usage constraints and downstream dependencies dynamically. Second, we leverage a \loweranchor completion as an anchor to drive bidirectional retrieval, enabling precise context integration without extensive fine-tuning. This allows for iterative refinement that enhances generation precision and repository coherence, addressing the limitations of existing methods that often overlook the orthogonal dimensions of function dependencies and usages.

\section{Conclusion}

In this paper, we present \approach, a novel repository-level code generation framework that enhances LLMs by inlining relevant upstream and downstream context from the code repository. By systematically integrating contextual information from both function callers and callees, \approach provides a richer, more natural understanding of the target function's environment within the specific repository. Extensive experiments on the \textsc{DevEval} and \textsc{REPOEXEC} datasets show that \approach consistently outperforms strong baselines across multiple metrics. Ablation studies and targeted analyses confirm that the core innovation of inlining—incorporating both upstream and downstream context—is the primary driver behind the significant improvements in return statement accuracy and function-call precision. Beyond these empirical gains, \approach demonstrates robust generalization across diverse programming domains and maintains stable performance when integrated with different backbone LLMs.

\section{Data Availability}

All code and data used in this study are publicly available at: \url{https://github.com/ythere-y/InlineCoder}.

\begin{acks}
    This research is funded by the National Key Research and Development Program of China (Grant No. 2023YFB4503802) and the Natural Science Foundation of Shanghai (Grant No. 25ZR1401175).
\end{acks}

\bibliographystyle{ACM-Reference-Format}
\bibliography{reference}

\end{document}

%% file: asserts/tables/total_dat_dev_eval.tex
\begin{table*}[t]
\centering
\small
\caption{Performance Comparison on the DevEval Dataset. The best performing baseline is underlined.}
\resizebox{\textwidth}{!}{
\begin{tabular}{l|cccc|cccc|cccc}
\toprule
\multirow{2}{*}{\textbf{Methods}}
  & \multicolumn{4}{c|}{DeepSeek-V3}
  & \multicolumn{4}{c|}{Qwen3-Coder}
  & \multicolumn{4}{c}{GPT5-mini} \\

\cmidrule(lr){2-5} \cmidrule(lr){6-9} \cmidrule(lr){10-13}
 & EM & ES & BLEU & ID.F1 
 & EM & ES & BLEU & ID.F1
 & EM & ES & BLEU & ID.F1 \\

\midrule

In-File &7.56 & 59.81 & 43.58 & 71.54 & 7.95 & 56.15 & 40.07 & 68.75 & 0.85 & 40.87 & 19.93 & 51.56 \\
Vanilla & \underline{11.45} & 66.20 & 52.64 & 76.99 & \underline{9.86} & \underline{60.02} & \underline{46.49} & \underline{74.92} & \underline{4.34} & 58.32 & 42.93 & \underline{64.43} \\
RepoCoder & 11.12 & \underline{68.61} & \underline{55.19} & \underline{77.26} & 2.90 & 45.00 & 29.09 & 72.57 & 3.65 & \underline{59.50} & \underline{43.72} & 62.40 \\
GraphCoder & 1.91 & 67.71 & 46.89 & 66.56 & 0.00 & 59.85 & 35.03 & 61.87 & 3.21 & 55.99 & 37.72 & 61.85 \\
DRACO & 0.64 & 47.72 & 28.06 & 46.17 & 0.00 & 51.91 & 31.25 & 41.12 & 1.27 & 42.93 & 23.60 & 45.19 \\
InlineCoder & \textbf{11.56} & \textbf{70.50} & \textbf{57.12} & \textbf{78.01} & \textbf{11.23} & \textbf{67.39} & \textbf{53.30} & \textbf{76.91} & \textbf{4.45} & \textbf{65.21} & \textbf{46.72} & \textbf{65.45} \\

\bottomrule
\end{tabular}
\label{tab:total_dev_eval}
}
\end{table*}

%% file: asserts/tables/total_data_repo_exec.tex
\begin{table*}[t]
\centering
\small
\caption{Performance Comparison on the REPOEXEC Dataset. The best performing baseline is underlined.}
\resizebox{\textwidth}{!}{
\begin{tabular}{l|cccc|cccc|cccc}
\toprule

\multirow{2}{*}{\textbf{Methods}}
  & \multicolumn{4}{c|}{DeepSeek-V3}
  & \multicolumn{4}{c|}{Qwen3-Coder}
  & \multicolumn{4}{c}{GPT5-mini} \\

\cmidrule(lr){2-5} \cmidrule(lr){6-9} \cmidrule(lr){10-13}
 & EM & ES & BLEU & ID.F1
 & EM & ES & BLEU & ID.F1
 & EM & ES & BLEU & ID.F1 \\

\midrule
In-File & 0.56 & 53.80 & 30.94 & 71.87 & 0.85 & 53.36 & 33.18 & 69.84 & 0.00 & 24.98 & 5.07 & 53.59 \\
Vanilla & \underline{0.85} & \underline{54.63} & \underline{31.57} & \underline{72.50} & 1.63 & 54.96 & 34.17 & 70.91 & 0.00 & 24.47 & 4.79 & 57.22 \\
RepoCoder & 0.28 & 53.01 & 30.43 & 72.08 & \underline{1.69} & \underline{56.75} & \underline{36.13} & \underline{71.17} & 0.00 & 33.87 & 13.33 & \underline{57.63} \\
GraphCoder & 0.00 & 39.69 & 14.37 & 48.65 & 0.00 & 30.68 & 9.00 & 51.98 & \underline{0.00} & \underline{40.19} & \underline{14.88} & 45.11 \\
DRACO & 0.28 & 52.98 & 29.25 & 47.29 & 0.00 & 55.27 & 33.72 & 48.08 & 0.00 & 33.55 & 12.28 & 51.83 \\
InlineCoder & \textbf{2.22} & \textbf{62.01} & \textbf{43.41} & \textbf{72.84} & \textbf{2.54} & \textbf{59.20} & \textbf{37.53} & \textbf{71.26} & \textbf{0.00} & \textbf{49.27} & \textbf{29.06} & \textbf{57.65} \\
\bottomrule
\end{tabular}
\label{tab:total_repo_exec}
}
\end{table*}

%% file: asserts/tables/ablation_dev_eval.tex
\begin{table*}[t]
\centering
\small
\caption{Ablation Study for Dataflow Analysis on the DevEval Dataset}
\label{tab:ablation_DevEval}
\begin{tabular}{l|cccc}
\toprule
\multirow{2}{*}{\textbf{Configuration}} & \multicolumn{4}{c}{DeepSeek-V3} \\
\cmidrule(lr){2-5}
& EM & ES & BLEU & ID.F1 \\
\midrule
InlineCoder & \textbf{11.56} & \textbf{70.50} & \textbf{57.12} & \textbf{78.01} \\
\;\;w/o upstream & 9.48 \textcolor{red!60}{(-2.12)} & 65.12 \textcolor{red!60}{(-5.38)} & 51.34 \textcolor{red!60}{(-5.78)} & 77.43 \textcolor{red!35}{(-0.58)} \\
\;\;w/o inline & 9.15 \textcolor{red!60}{(-2.41)} & 65.95 \textcolor{red!60}{(-4.55)} & 52.35 \textcolor{red!60}{(-4.77)} & 78.01 \textcolor{red!35}{(0.00)} \\
\;\;w/o downstream & 9.75 \textcolor{red!35}{(-1.81)} & 65.87 \textcolor{red!60}{(-4.63)} & 52.25 \textcolor{red!60}{(-4.87)} & 77.60 \textcolor{red!35}{(-0.41)} \\
\;\;w/o confidence & 10.74 \textcolor{red!35}{(-0.82)} & 67.79 \textcolor{red!35}{(-2.71)} & 54.24 \textcolor{red!35}{(-2.88)} & 77.30 \textcolor{red!60}{(-0.71)} \\
\;\;w/o draft & 7.67 \textcolor{red!100}{(-3.89)} & 59.09 \textcolor{red!100}{(-11.41)} & 44.24 \textcolor{red!100}{(-12.88)} & 76.29 \textcolor{red!100}{(-1.72)} \\ 
\bottomrule
\end{tabular}
\end{table*}

%% file: asserts/tables/validation_upstream.tex
\begin{table*}[t]
\centering
\small
\caption{Comparison of the Last Line of the Return Statement} 
\begin{tabular}{l|c|c|c}
\toprule
\textbf{Methods} & EM & BLEU   & ES \\ 
\midrule
Vanilla      & 35.40      & 50.17          & 58.72  \\ 
RepoCoder & 37.15 & 52.77 & 61.57 \\ 
InlineCoder     & \textbf{37.92}       & \textbf{53.80}         & \textbf{62.51} \\ 
\bottomrule
\end{tabular}
\label{tab:validation_upstream}
\end{table*}

%% file: asserts/tables/validation_downstream.tex
\begin{table*}[t]
\centering
\small
\caption{Comparison of Call Statements}
\begin{tabular}{l|c|c|c|c|c}
\toprule
\textbf{Methods} & EM & Jaccard  & F1  & Coverage & DIR \\ 
\midrule

Vanilla      & 29.86      & 53.10          & 60.67  &63.16 & 69.21\\ 
RepoCoder   & 21.48 &52.80 & 63.01 & 65.13 & 70.45 \\
InlineCoder     & \textbf{30.74}        & \textbf{54.99}         & \textbf{62.88} & \textbf{65.40} &\textbf{71.86}\\ 

\bottomrule
\end{tabular}
\label{tab:validation_downstream}
\end{table*}